\documentstyle[preprint,aps]{revtex}
\begin{document}
\draft

\title{Probing Quark-Gluon Correlation Functions}

\author{Xiaofeng Guo$^1$ and Jianwei Qiu$^{2,3}$}
\address{$^1$Department of Physics and Astronomy, 
             University of Kentucky,\\
             Lexington, Kentucky 40506, USA\\
         $^2$Department of Physics and Astronomy, 
             Iowa State University \\
             Ames, Iowa 50011, USA \\
         $^3$Physics Department, 
             Brookhaven National Laboratory\\
             Upton, New York 11973-5000, USA }
\date{November 2, 1999}
\maketitle

\begin{abstract}
We review applicability of QCD factorization theorem to multiple
scattering in deeply inelastic lepton-nucleus scattering.  We show
why $A^{1/3}$-type nuclear enhancement can be calculated consistently
in perturbative QCD.  We derive the transverse momentum broadening of
the leading pions in deeply inelastic lepton-nucleus collisions.  We
argue that the measurement of such transverse momentum broadening can
provide direct information on the multiple-parton correlation
functions inside a nucleus.  We also estimate the numerical values of
the broadening at different values of $x_B$ and $Q^2$. 
\end{abstract}
\vspace{0.2in}

\section{Introduction}

Multiparton correlation functions inside a large nucleus
are extremely important and useful for understanding nuclear
dependence in relativistic heavy ion collisions.   
When a quark or a gluon passes through the nuclear matter, it
loses energy via radiation, and it picks up extra transverse 
momentum because of multiple scattering
\cite{Mueller,GW,AHMetal,BBL,LQS1}.  
The multiple scattering is directly associated with 
multiparton correlation functions \cite{LQS1}. 
Inside a large nucleus, multiple scattering can take
place within one nucleon or between different nucleons.  At high
energy, a single hard scattering is always localized
within one nucleon, and it cannot generate any large dependence on 
the nuclear size.  Similarly, multiple scattering within one nucleon
cannot provide much dependence on the nuclear size either.  Therefore,
the large dependence on the nuclear size is an unique signal of multiple
scattering between nucleons.  Measurements of such anomalous 
dependence on the nuclear size for any physical observable
will provide direct information on multiparton correlation 
functions in a nucleus.  

However, there are many multiparton correlation functions in 
QCD \cite{LQS1}.  It is therefore important to identify physical
observables 
which depend only on a small number of multiparton correlation
functions. Otherwise, data will not be able to separate
contributions from different correlation functions.  It was pointed
out recently in Ref.~\cite{Guo2} that 
the $A^{1/3}$-type enhancement of the transverse momentum
broadening of Drell-Yan pairs and the jet broadening in 
lepton-nucleus deeply inelastic scattering (DIS) are good observables, 
because at the leading order of $\alpha_s$, these two observables
depend only on one type of multiparton correlation functions.  It
represents the correlations between hard quarks and soft-gluons,
and has the following operator definition \cite{LQS1,Guo2},
\begin{eqnarray}
T_{qF}^{A}(x,\mu^2)&=& \int \frac{dy^-}{2\pi} 
e^{ix p^+ y^-} \frac{dy_1^- dy_2^-}{2\pi}
\theta(y_1^--y^-)\theta(y_2^-)
\nonumber \\
&\times &
\langle P_A| \bar{\psi}_q(0)\, \frac{\gamma^+}{2}\, 
F^+_{\ \alpha}(y_2^-) \, F^{+\alpha}(y_1^-)\, 
\psi_q(y^-)\, |P_A\rangle \, ,
\label{T4qA}
\end{eqnarray}
with $A$ the atomic weight, and $q$ is the quark flavor and $F$ is the
gluon field strength.  In Eq.~(\ref{T4qA}), $x$ is the effective
momentum fraction and $\mu$ represents the scale-dependence of the
correlation functions.  

According to QCD factorization theorem \cite{CSS_fac,QS_fac}, all
information on the identity of the target is factorized into the
matrix elements.  The $A^{1/3}$-type enhancement must be a property of
the relevant nuclear matrix elements, like the one in
Eq.~(\ref{T4qA}).  Experimentally, the $A^{1/3}$-type nuclear
enhancement have been observed in the transverse momentum 
broadening of Drell-Yan pairs in hadron-nucleus collisions
\cite{DY_PT}.  Theoretically, Eq.~(\ref{T4qA}) shows how such
enhancement can occur, through integrals over the relative positions
of the fields, $y_i$, that appear in the expectation values.
For example, enhancement can occur when the two quark
fields are close together and the two gluon fields are close together, 
and they pair off into color singlets that are separated by no more
than a nucleon diameter in the $y_i$ integrals; but the two pairs are
separated by a distance that varies up to the nuclear size.  In this
manner, aside from an overall factor of $A$ which reflects growth with
the nuclear volume, we anticipate an $A^{1/3}$ nuclear enhancement 
\cite{LQS2}.  

Direct measurement of the multiparton correlation functions is of
great importance for testing the generalized QCD factorization
theorem \cite{QS_fac}, which allows us to 
apply QCD perturbation theory for studying the collisions involving
nuclei.  It was observed in Ref.~\cite{Guo2} that 
measuring the jet transverse momentum broadening in DIS,
$\Delta\langle p_T^2\rangle$, provides a {\it direct} measurement of
the quark-gluon correlation functions, $T^A_{qF}(x_B,Q^2)$, inside a
nucleus.  At the lowest order, the jet transverse momentum broadening
in DIS can be expressed as 
\cite{Guo2} 
\begin{eqnarray}
\Delta\langle p_T^2\rangle &\equiv & 
\langle p_T^2\rangle^{eA} - \langle p_T^2\rangle^{eN} 
\nonumber \\
&\approx & \frac{4\pi^2\alpha_s(Q^2)}{3}\,
    \frac{\sum_q e_q^2\,T_{qF}^{A}(x_B,Q^2)}
         {\sum_q e_q^2\,q^{A}(x_B,Q^2)}\, ,
\label{jetdis}
\end{eqnarray}
where $p_T$ is the transverse component of the jet momentum in the
photon-nucleus center of mass frame in DIS.  In Eq.~(\ref{jetdis}),
$x_B$ is the  
Bjorken variable and $Q^2=-q^2$ with $q^2$ is the invariant mass of
the virtual photon in DIS.  

However, other than a future HERA with a nuclear beam, existing
fixed target facilities can not provide good measurements on jets in
lepton-nucleus DIS. It is the purpose of this paper to show that by
measuring the leading pions in DIS and their transverse momentum
broadening, fixed target facilities, such as CEBAF \cite{Brodsky}, 
can provide good information on the quark-gluon correlation functions 
defined in Eq.~(\ref{T4qA}).  

This paper is organized as follows.  In the next section, we derive the
transverse momentum broadening of the leading pions in lepton-nucleus
DIS.  We demonstrate analytically the direct correspondence between
the transverse momentum broadening of the leading pions and 
multiparton correlation functions.  In Sec.~III, we estimate the
numerical values of such broadening at different values of $x_B$ and
$Q^2$, by using the model of the quark-gluon correlation functions
introduced in Ref.~\cite{LQS2}. We also derive explicit relations of
the transverse momentum broadening for $\pi^{\pm}$ and $\pi^0$,
and discuss the $x_B$ and $Q^2$ dependence of the correlation
functions.


\section{Transverse momentum broadening of pions in DIS}

Consider leading pion production in the deeply inelastic 
lepton-nucleus scattering, 
\begin{equation}
e(k_1) + A(p) \longrightarrow e(k_2) + \pi(\ell) +X \ ,
\label{process}
\end{equation}
where $k_1$ and $k_2$ are the four
momenta of the incoming and the outgoing leptons respectively,  
$p$ is the momentum per nucleon for the nucleus with the atomic 
number $A$, and $\ell$ is the observed pion momentum.  
In order to extract useful information on multiple scattering from
the pion production, it is natural to think that the $A$-dependence of
the differential cross section $d\sigma_{eA\rightarrow e\pi X}
/dx_BdQ^2d\ell_T^2$ is the physical observable to study, 
because the transverse momentum spectrum is more sensitive to 
multiple scattering \cite{LQS1}.  But, other than a future HERA with
a nuclear beam, existing fixed target facilities can not produce good
data on the production of leading pions at {\it large} transverse
momenta in DIS.  On the other hand, at the small transverse momentum 
$\ell_T^2$, the differential cross section $d\sigma/dx_BdQ^2d\ell_T^2$
is proportional to $1/\ell_T^2$, and is not perturbatively stable.    
A nontrivial all order resummation of the large logarithms
$\log(Q^2/\ell_T^2)$ is necessary in order to get a reliable
prediction of the transverse momentum spectrum at small $\ell_T^2$
\cite{Guo1}.  However, the inclusive moments of the transverse
momentum spectrum $\int d\ell_T^2 (\ell_T^2)^n\, 
d\sigma_{eA\rightarrow e\pi X}/dx_BdQ^2d\ell_T^2$ with $n\ge 1$ are
perturbatively stable, and also enhance the information in the region of
large $\ell_T^2$, where the multiple scattering is most relevant.
Therefore, the $A$-dependence for the moments of the transverse momentum
spectrum of the leading pions is a good observable for extracting
information on multiple scattering.

We define the first moment of the transverse momentum spectrum - the
averaged pion transverse momentum squared, as 
\begin{equation}
\langle \ell_{T}^2\rangle^{eA} = \left.
\int d\ell_{T}^2 \, \ell_{T}^2 \, 
\frac{d\sigma_{eA\rightarrow e\pi X}}{dx_B  dQ^2 d\ell_{T}^2} 
\right/ \frac{d\sigma_{eA\rightarrow eX}}{dx_B  dQ^2} \, ;
\label{LTavg}
\end{equation}
where $d\sigma_{eA\rightarrow eX}/dx_B dQ^2$ is the total inclusive
DIS cross section normalized by the atomic weight $A$.  
In Eq.~(\ref{LTavg}), the Bjorken variable
$x_B=Q^2/(2p\cdot q)$, and $q=k_1-k_2$ is the four-momentum of the
virtual photon.  The transverse momentum $\ell_{T}$ of the leading
pion depends on our choice of the frame.  In this paper, we choose the
photon-nucleus frame, and choose the $z$-axis along the momentum line
of the nucleus and the virtual photon. 

To separate the multiple scattering contribution from 
the single scattering contribution, we define the nuclear 
broadening of the transverse momentum square as
\begin{equation}
\Delta\langle \ell_T^2\rangle \equiv  
\langle \ell_T^2\rangle^{eA} - \langle \ell_T^2\rangle^{eN} \, .
\label{DLT2}
\end{equation}
As we will demonstrate below, the nuclear broadening of the transverse
momentum square defined in Eq.~(\ref{DLT2}) can be parameterized as 
\cite{E683,E609}
\begin{equation}
\Delta\langle \ell_T^2\rangle =
a + b\, A^{1/3}\ ,
\label{BA13}
\end{equation}
where $bA^{1/3}$ represents
the contribution directly from the multiple scattering which is
explicitly proportional to the nuclear size ($\propto A^{1/3}$) 
and term $a$ includes all contributions from the {\it localized} hard
scattering as well as those suppressed by high power of $1/Q^2$.  
In principle, the parameter $a$ in Eq.~(\ref{BA13}) can also depend on
the atomic weight $A$. 
But, as we will explain below, its dependence on $A$ should be
very weak (e.g., $A^{\alpha}$ dependence with $\alpha \simeq 
\pm 0.02$).  In this paper, we are interested in the second term in
Eq.~(\ref{BA13}), and we show that experimental measurement of 
the constant $b$ in Eq.~(\ref{BA13}) 
will provide direct information on the quark-gluon correlation
functions shown in Eq.~(\ref{T4qA}).

To derive explicit expressions for the $a$ and $b$ in
Eq.~(\ref{BA13}), we expand both numerator and the denominator in
Eq.~(\ref{LTavg}) in terms of a power series of $1/Q^2$.  In addition
to the terms of leading power, we keep only the power suppressed terms
that are explicitly enhanced by a factor $A^{1/3}$.  Since the
denominator, $d\sigma_{eA\rightarrow eX}/dx_B dQ^2$, is the total
inclusive DIS cross section, the operator product expansion (OPE)
allows us to expand it in terms of the power series of $1/Q^2$
\cite{CFP,Ellis},
\begin{eqnarray}
\frac{d\sigma_{eA\rightarrow eX}}{dx_B dQ^2} &=& 
\sum_a \int dx\, \phi_{a/A}(x,\mu^2)\,
       \frac{d\hat{\sigma}^{(0)}_{ea\rightarrow eX}}{dx_B dQ^2}
       \left(x_B/x,\mu^2/Q^2,\alpha_s(\mu^2)\right)
\left[\, 1 + O\left(\frac{1}{Q^2}\right)\right]\ 
\nonumber \\
&\equiv & D_A^{(0)}\left[\, 1 + O\left(\frac{1}{Q^2}\right)\right]\ , 
\label{D0}
\end{eqnarray}
where $\mu$ is the factorization and/or renormalization scale, 
$\phi_{a/A}(x,\mu^2)$ is the leading twist parton distribution
of flavor $a$ in a nucleus normalized by the atomic weight $A$, and  
$d\hat{\sigma}^{(0)}_{ea\rightarrow eX}/dx_B dQ^2$ is a
perturbatively calculable short-distance hard part, which is
independent of the nuclear medium.  The {\it total}\ DIS
cross section normalized by the atomic weight $A$ is an inclusive
quantity and depends only on one hard scale $Q^2$.  The QCD 
factorization theorem \cite{CFP,Ellis} allows us to factorize each
term in Eq.~(\ref{D0}) into a convolution of a short-distance
coefficient function and a corresponding matrix element. The
$1/Q^2$ term in Eq.~(\ref{D0}) can be expressed in terms of twist-4
matrix elements \cite{Ellis}.  As demonstrated in
Refs.~\cite{QS_fac,Qiu}, integration of all position variables
($y_i$) of the field operators, which define these twist-4 matrix
elements, are bounded by $1/x_Bp$ due to the oscillating exponential.  
Let $m$ and $r$ be nucleon mass and radius, respectively. 
If $1/x_Bp < 2r(m/p)$ (i.e. $x_B > 1/2mr \approx 0.1$), all the fields
in these twist-4 matrix elements are bounded within the size of
individual nucleon.  Therefore, the power suppressed terms in
Eq.~(\ref{D0}) is of the order of $O(1/Q^2)$ if $x_B > 0.1$, not
$O(A^{1/3}/Q^2)$.  

On the other hand, the numerator in Eq.~(\ref{LTavg}) depends on two
physically measured hadrons: the pion and the nucleus, and therefore,
the OPE alone cannot be used to expand the numerator.  However, the
generalized factorization theorem introduced in Ref.~\cite{QS_fac} can
be used to factorize the numerator up to the $1/Q^2$ power
corrections.  Similarly to the Eq.~(1) of Ref.~\cite{LQS2}, we expand
the numerator in Eq.~(\ref{LTavg}) as
\begin{eqnarray}
\int d\ell_{T}^2 \, \ell_{T}^2 \, 
\frac{d\sigma_{eA\rightarrow e\pi X}}{dx_B  dQ^2 d\ell_{T}^2} 
&=&
\sum_{a,c} \phi_{a/A}(x) \otimes 
   C^{(0)}_{ea\rightarrow ecX}\left(x_B/x,z=\ell/p_c,Q^2\right)
           \otimes D_{c\rightarrow\pi}(z)
\nonumber \\
&+&\frac{1}{Q^2}
\sum_{a,c} \left[
           T_{a/A}(x) \otimes 
   C^{(2)}_{ea\rightarrow ecX}\left(x_B/x,z=\ell/p_c,Q^2\right)
           \otimes D_{c\rightarrow\pi}(z) \right.
\nonumber \\ 
&& {\hskip 0.4in} \left.
          +\phi_{a/A}(x) \otimes 
\bar{C}^{(2)}_{ea\rightarrow ecX}\left(x_B/x,z=\ell/p_c,Q^2\right)
           \otimes d_{c\rightarrow\pi}(z) \right]
\nonumber \\
&+& ... 
\nonumber \\
&\equiv &
H^{(0)}_A + H^{(2)}_A + \bar{H}^{(2)}_A 
+ ...\, ,
\label{Hexp}
\end{eqnarray}
where $...$ represents terms further suppressed in $1/Q^2$, $\otimes $
represents the convolution over partons' momentum 
fractions, and explicit dependence on the factorization and/or
renormalization scale is suppressed for simplicity.  In
Eq.~(\ref{Hexp}), $C^{(0)}$, $C^{(2)}$, and $\bar{C}^{(2)}$ are
perturbatively calculable short-distance hard parts \cite{QS_fac}, and
$D_{c\rightarrow \pi}$ and $d_{c\rightarrow \pi}$ are twist-2 and
twist-4 parton-to-pion fragmentation functions, respectively.  
$\phi_{a/A}$ are nuclear parton distributions, and 
the $T_{a/A}$ in Eq.~(\ref{Hexp}) represents the twist-4 quark-gluon
correlation functions (e.g., the one defined in Eq.~(\ref{T4qA})).
Both $\phi_{a/A}$ and $T_{a/A}$ are normalized by the atomic weight
$A$.

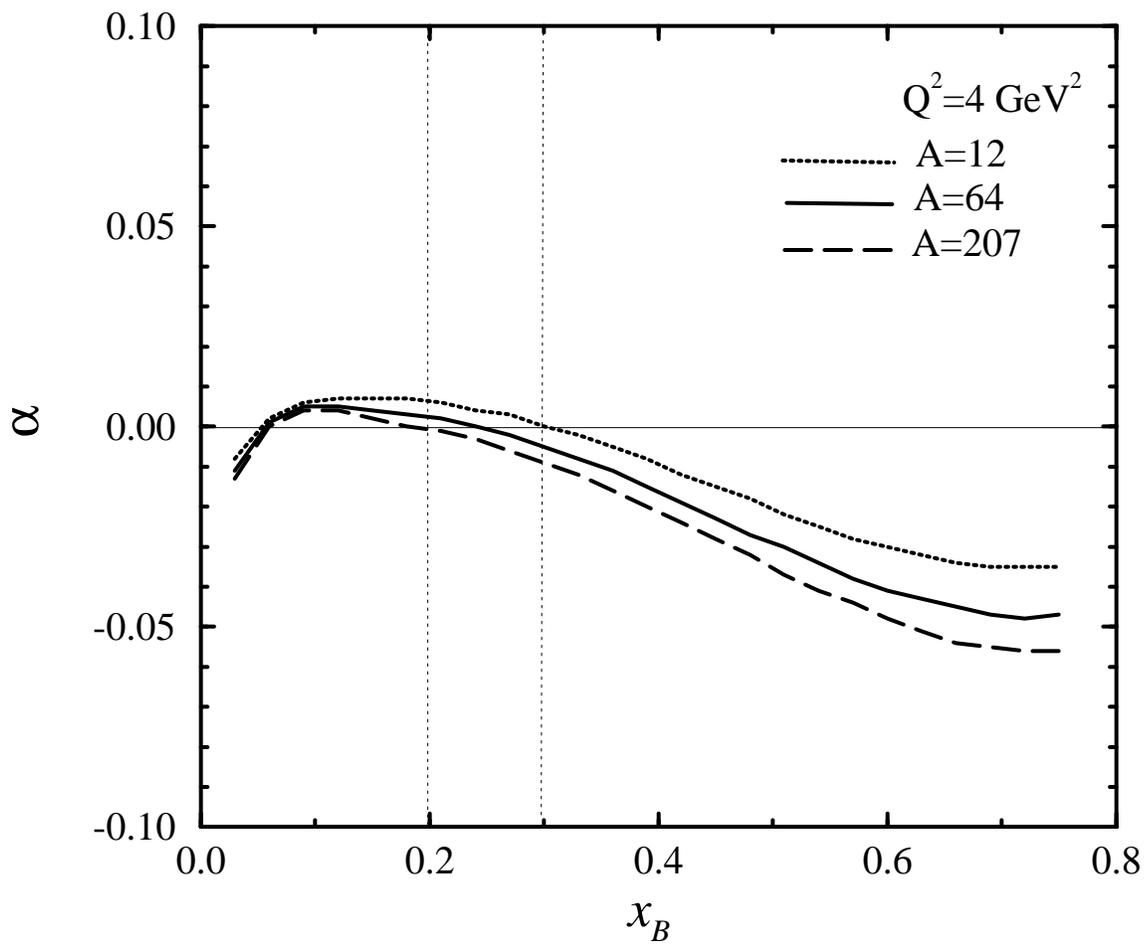
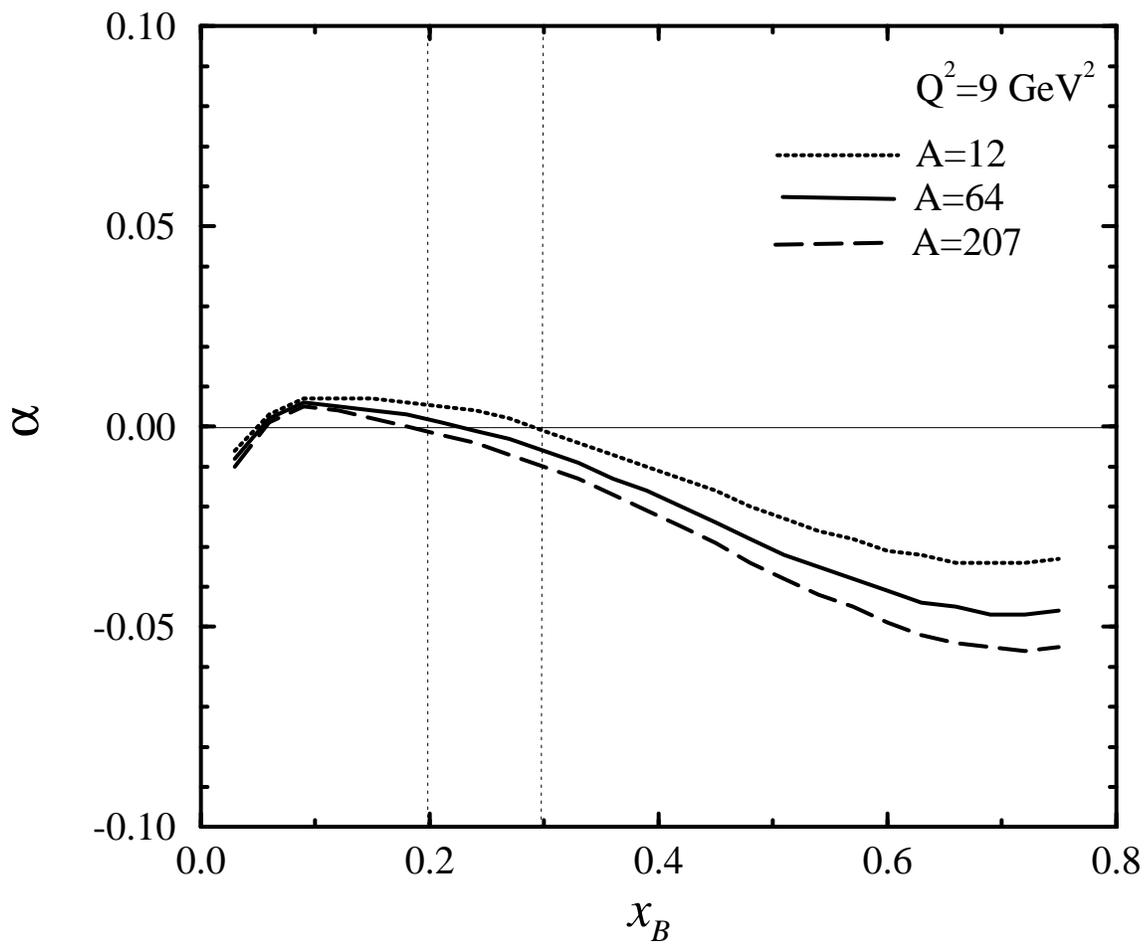
\begin{figure}
\caption{ The $\alpha(A,x_B)$ defined in Eq.~(\protect{\ref{F2}}) as a
function of $x_B$ at (a) $Q^2=4$~GeV$^2$ and (b) $Q^2=9$~GeV$^2$.}
\label{fig0}
\end{figure}

Effective nuclear parton distributions $\phi_{a/A}(x)$
have been measured \cite{EMC,SLAC,NMC,E665} and are known to have the
nuclear shadowing for small $x$ ($x\simeq 0.1$), the EMC effect for
intermediate $x$ values ($0.3\simeq x \simeq 0.7$), and Fermi motion
effect for the large $x$ region.  By fitting recent high precision
DIS and Drell-Yan data on various nuclear targets, Eskola {\it et al.}
\cite{Eskola} extracted a set of effective scale dependent nuclear
parton distributions.  In order to estimate the $A$-dependence of
the nuclear parton distributions, we introduce a parameter
$\alpha(A,x_B)$ as 
\begin{equation}
F_2^A(x_B,Q^2) \equiv A^{\alpha(A,x_B)}\, F_2^N(x_B,Q^2)\ ,
\quad\quad
\mbox{or}
\quad\quad
\alpha(A,x_B) =
\frac{\ln\left[F_2^A(x_B,Q^2)\left/F_2^N(x_B,Q^2)\right.\right]}
     {\ln\left[A\right]}\, ,
\label{F2}
\end{equation}
where $F_2^A(x_B,Q^2)$ is a nuclear structure function normalized by
$A$ and $F_2^N(x_B,Q^2)$ is an isoscalar nucleon structure function.
Using the lowest order expression for the structure function: 
$F_2(x_B,Q^2)=\sum_q\, e_q^2\, x_B\, \phi_{q}(x_B,Q^2)$ with $e_q$
being a quark's fractional charge and $\phi_{q}$ the effective nuclear 
parton distributions of Ref.~\cite{Eskola}, we plot the parameter
$\alpha(A,x_B)$ as a function of $x_B$ for $A$=12, 64, and 
207 at $Q^2=4$ and 9 GeV$^2$ in Fig.~\ref{fig0}a and \ref{fig0}b,
respectively.  Fig.~\ref{fig0} shows that although the exact
$x_B$-dependence and the $A$-dependence of the structure functions are
non-trivial, the overall $A$-dependence of the structure functions is
limited to $A^{\alpha}$ with a value of $\alpha\simeq \pm 0.02$ for
$x_B$ values between two vertical lines in Fig.~\ref{fig0}, which are
relevant to pion production in this paper. 
Actually, as shown in Fig.~\ref{fig0}, the absolute value of $\alpha$
is limited to 0.05 for any practical value of $x_B$ accessible at  
fixed target energies.
Such $A$-dependence in the structure function as well as in the
$\phi_{a/A}(x)$ is much weaker than the 
$A^{1/3}$-dependence caused by multiple scattering.
Since the parton-to-pion fragmentation functions, $D_{c\rightarrow
\pi}$ and $d_{c\rightarrow \pi}$, should not have explicit
$A^{1/3}$-dependence, the term $\bar{H}^{(2)}_A$ is of the order of 
$O(1/Q^2)$, not $O(A^{1/3}/Q^2)$.  On the other hand, the term
$H^{(2)}_A$ depends on the multi-parton correlation function in a
nucleus $T_{a/A}(x)$, and therefore, it will have an $A^{1/3}$-type
enhancement \cite{LQS2}.

Substituting Eqs.~(\ref{D0}) and (\ref{Hexp}) into Eq.~(\ref{LTavg}),
and keeping terms up to $O(A^{1/3}/Q^2)$, we obtain
\begin{equation}
\langle \ell_{T}^2\rangle^{eA} \approx
\frac{H^{(0)}_A+H^{(2)}_A}{D^{(0)}_A} +
O\left(\frac{A^0}{Q^2}\right)\ .
\label{expLTA}
\end{equation}
Similarly, for a nucleon target, we have 
\begin{equation}
\langle \ell_{T}^2\rangle^{eN} \approx
\frac{H^{(0)}_N}{D^{(0)}_N} + 
O\left(\frac{A^0}{Q^2}\right)\ .
\label{expLTN}
\end{equation}
Substituting above Eqs.~(\ref{expLTA}) and (\ref{expLTN}) into our
definition of the nuclear broadening of the transverse momentum square
in Eq.~(\ref{DLT2}), we derive
\begin{equation}
\Delta\langle \ell_T^2\rangle \approx 
\left[\frac{H^{(0)}_A}{D^{(0)}_A} -
      \frac{H^{(0)}_N}{D^{(0)}_N} \right] +  
\frac{H^{(2)}_A}{D^{(0)}_A} + 
O\left(\frac{A^0}{Q^2}\right)\ ,
\label{DLT2com}
\end{equation}
where the power suppressed terms $O(A^0/Q^2)$ should be very small due 
to the cancellation between $\langle \ell_{T}^2\rangle^{eA}$ and 
$\langle \ell_{T}^2\rangle^{eN}$ and the fact that they are not
enhanced by the $A^{1/3}$.  The first term in Eq.~(\ref{DLT2com})
would be exactly equal to zero if the normalized effective 
nuclear parton distributions $\phi_{a/A}(x)=\phi_{a/N}(x)$.  However,
due to the well-known nuclear effects in the parton distributions, the
first term in Eq.~(\ref{DLT2com}) does not have to vanish.  But, 
because of very weak $A$-dependence of nuclear parton distributions
shown in Fig.~\ref{fig0}, we identify
this term with the $a$ in Eq.~(\ref{BA13}),
\begin{equation}
a \approx \frac{H^{(0)}_A}{D^{(0)}_A} -
      \frac{H^{(0)}_N}{D^{(0)}_N}\ .
\label{DLTa}
\end{equation}
It is the second term in Eq.~(\ref{DLT2com})
that is responsible for the $A^{1/3}$-type dependence of the nuclear
broadening.  

We introduce $\Delta \langle \ell_T^2\rangle_{1/3}$ as  
\begin{eqnarray}
\Delta \langle \ell_T^2\rangle_{1/3} &\equiv & b\, A^{1/3} 
\approx \frac{H^{(2)}_A}{D^{(0)}_A} 
\nonumber \\
&=& \left. 
\int d\ell_T^2 \, \ell_T^2 \, 
\frac{d\sigma^{(D)}_{eA\rightarrow e\pi X}}{dx_B dQ^2 d\ell_T^2} 
\right/ \frac{d\sigma^{(0)}_{eA\rightarrow eX}}{dx_B  dQ^2} \, ,
\label{dlT20}
\end{eqnarray}
where superscript ``(D)'' represents the double scattering
contribution, and ``(0)'' stands for the leading power contribution.
Our formalism in Eqs.~(\ref{DLTa}) and (\ref{dlT20}) can be verified
from the $A$-dependence of the measured transverse momentum
broadening.   

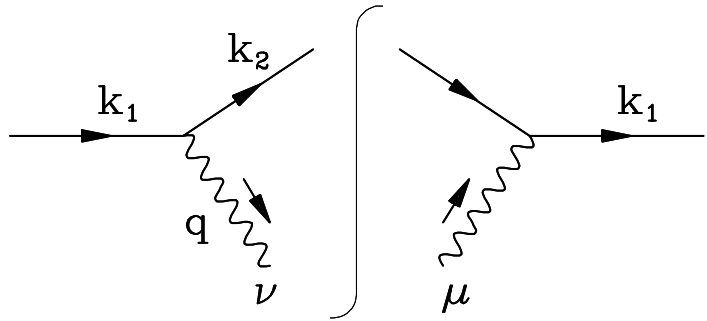
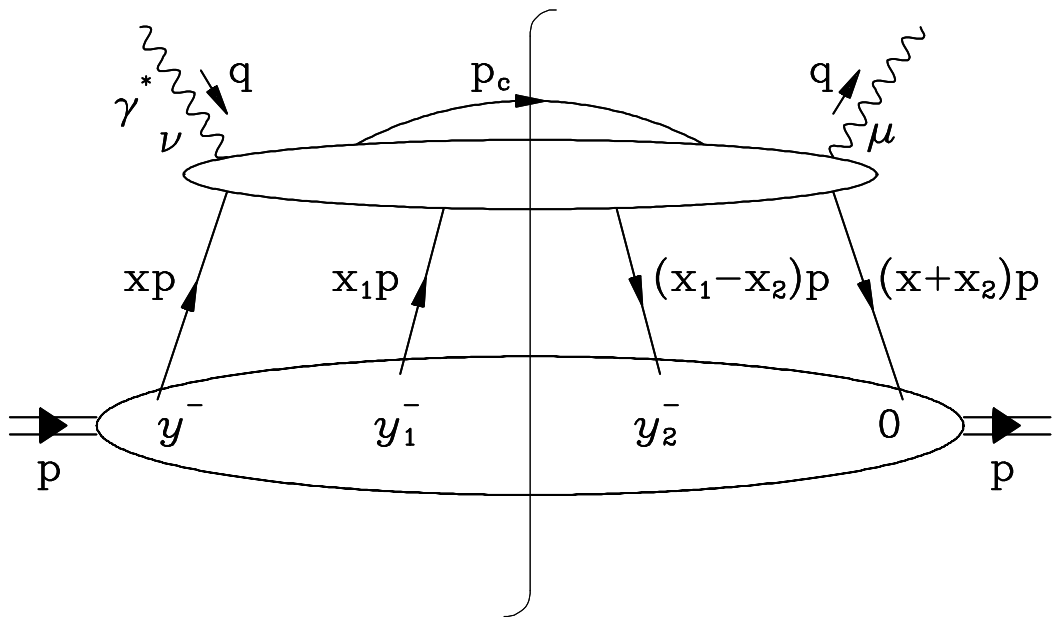
\begin{figure}
\caption{ (a) Diagram representing $L^{\mu\nu}$; 
(b) Diagram representing $W^{(D)}_{\mu\nu}$.}
\label{fig1}
\end{figure}

From the factorized formula in Eq.~(\ref{Hexp}), the double scattering
contribution $H_A^{(2)}$ depends only on the twist-2 parton-to-pion
fragmentation function $D(z)$.  Therefore, the differential cross
section in Eq.~(\ref{dlT20}) for the pion production in
DIS can be expressed as
\begin{equation}
\frac{d\sigma^{(D)}_{eA\rightarrow e\pi X}}{dx_B  dQ^2 d\ell_T^2}
=\sum_{c}\, \int 
\frac{d{\sigma}^{(D)}_{eA\rightarrow ecX}}{dx_B dQ^2 dp_{c_T}^2}
\cdot D_{c \rightarrow \pi} (z) \cdot \frac{dz}{z^2} \ ,
\label{sigma-pi}
\end{equation}
where $\sum_c$ runs over all parton flavors $c$, $z=\ell/p_c$ is the
momentum fraction carried by the produced pion, and
$D_{c\rightarrow\pi}(z)$ is the fragmentation function for 
the parton of momentum $p_c$ to become a pion of momentum $\ell$.   
In Eq.~(\ref{sigma-pi}), $d{\sigma}^{(D)}_{eA\rightarrow ecX}/
dx_B dQ^2 dp_{c_T}^2$ is the double scattering contributions to the
differential cross section to produce a parton of momentum $p_c$ in
DIS, and it can be written as 
\begin{equation}
\left[
\frac{d{\sigma}^{(D)}_{eA\rightarrow ecX}}{dx_B dQ^2 dp_{c_T}^2}
\right]\, dp_{c_T}^2
=\frac{1}{8\pi}\, \frac{e^2}{x_B^2s^2Q^2} \,
L^{\mu\nu}(k_1,k_2)\, W^{(D)}_{\mu\nu}(x_B,Q^2,p_c)\ ,
\label{sigma-c}
\end{equation}
where $s=(p+k_1)^2$ is the total invariant mass of the lepton-nucleon
system.
In Eq.~(\ref{sigma-c}), the leptonic tensor $L^{\mu\nu}$ is 
given by the diagram in Fig.~\ref{fig1}a,  
\begin{equation}
L^{\mu\nu}(k_1,k_2)
=\frac{1}{2}\, {\rm Tr}(\gamma \cdot k_1 \gamma^{\mu}
\gamma \cdot k_2 \gamma^{\nu}) \ ,
\label{L}
\end{equation}
and $W^{(D)}_{\mu\nu}$ is the hadronic tensor due to double
scattering, which is given by the diagram 
shown in Fig.~\ref{fig1}b.  For comparison, the lowest order
differential cross section for producing a parton of momentum $p_c$
due to single scattering is given by
\begin{equation}
\left[
\frac{d{\sigma}^{(0)}_{eA\rightarrow ecX}}{dx_B dQ^2 dp_{c_T}^2}
\right]\,
dp_{c_T}^2
= \left[
\frac{d{\sigma}^{(0)}_{eA\rightarrow ecX}}{dx_B dQ^2}\, 
  \delta(p_{c_T}^2) \right]\,
dp_{c_T}^2
\label{single}
\end{equation}
where $d{\sigma}^{(0)}/dx_B dQ^2$ is the leading order inclusive DIS
cross section, which appears in the definition of
$\Delta\langle\ell^2_T\rangle_{1/3}$ in Eq.~(\ref{dlT20}).

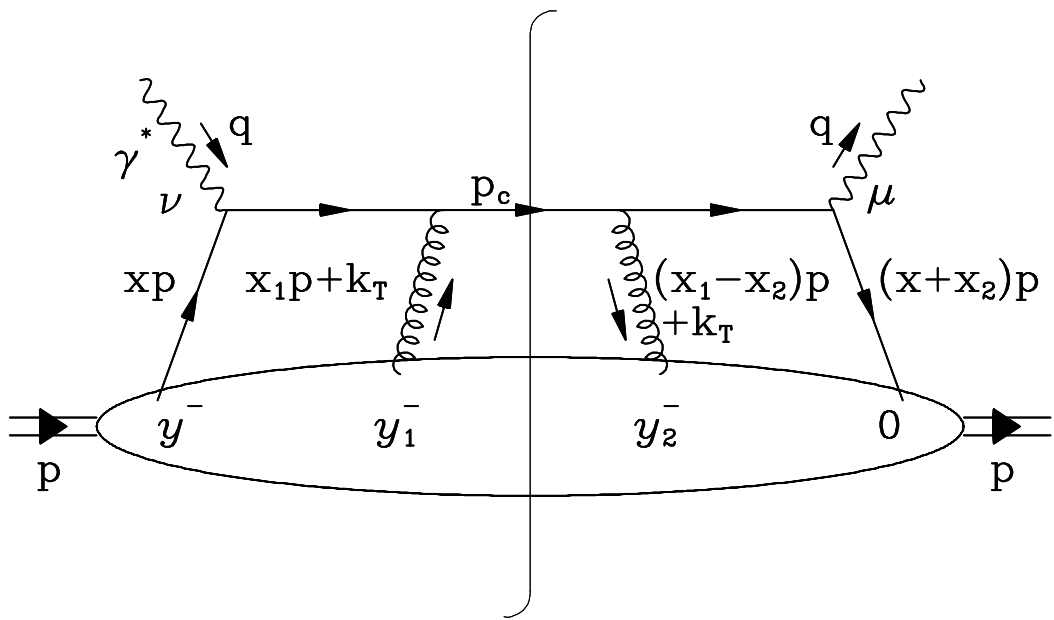
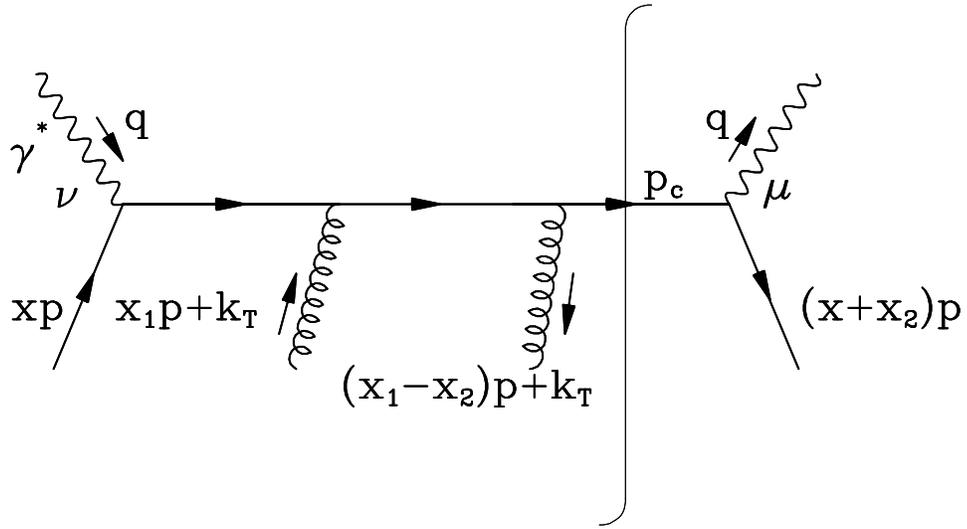
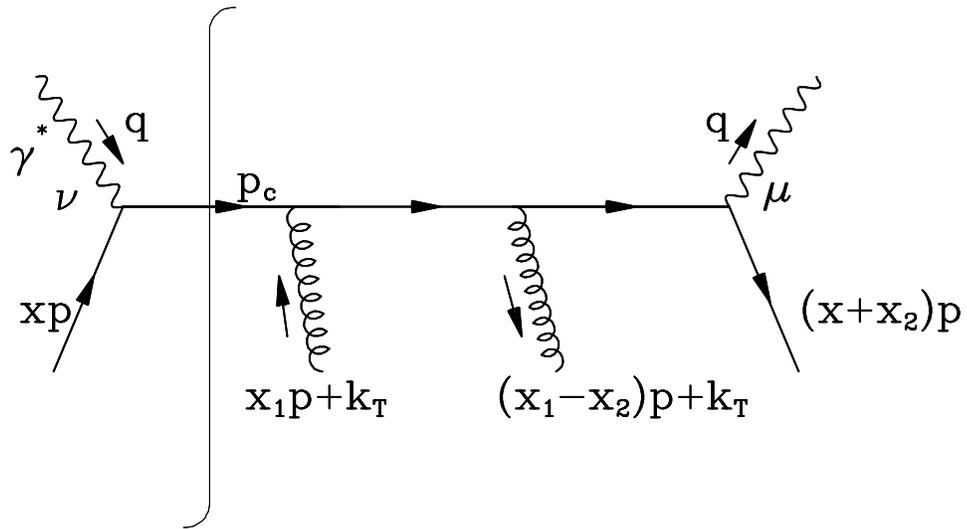
\begin{figure}
\caption{Lowest order double scattering contribution to the broadening
of the parton $c$: 
(a) symmetric diagram; (b) and (c): interference diagrams.}
\label{fig2}
\end{figure}

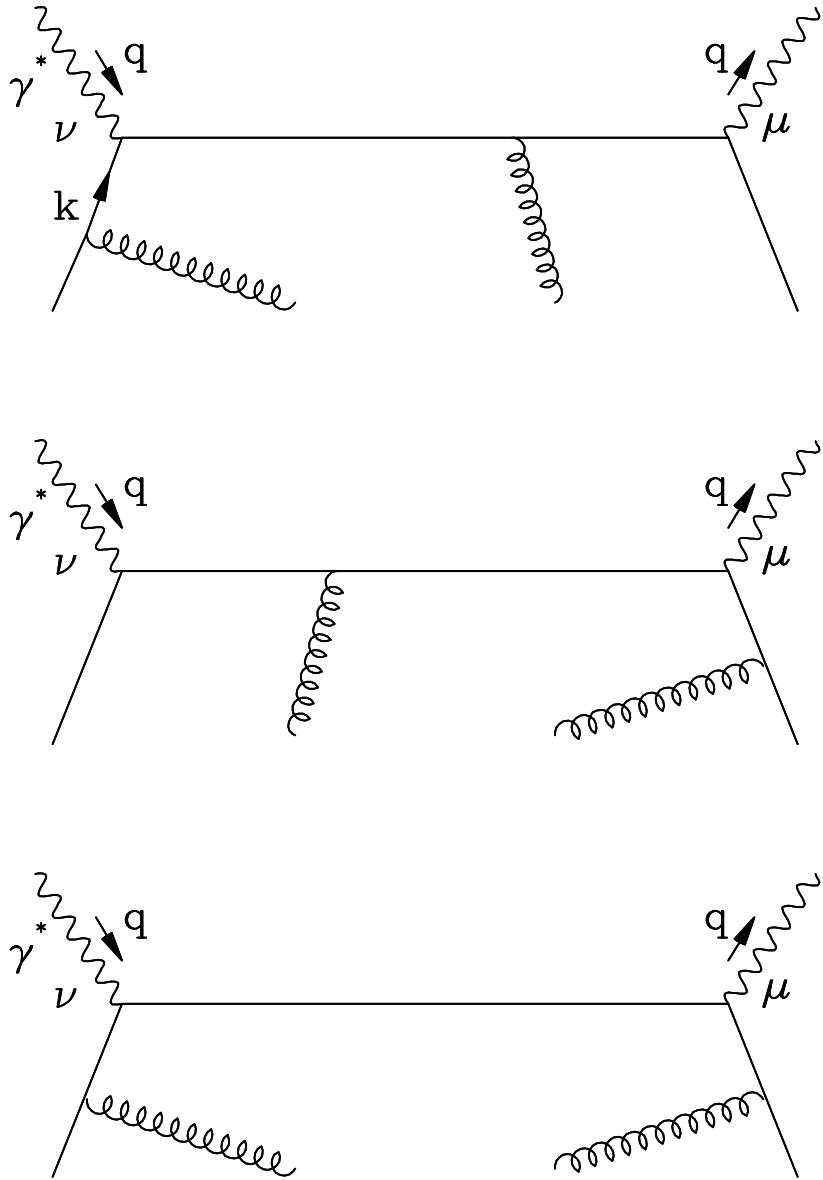
\begin{figure}
\caption{Lowest order double scattering diagrams which do not
contribute to the broadening of the parton $c$.} 
\label{fig2p}
\end{figure}

The complete double scattering contributions to the production of
a leading quark of momentum $p_c$ in DIS at the leading order of
$\alpha_s$ come from the diagrams shown in Fig.~\ref{fig2}, which 
represent the final-state interactions, plus the same order diagrams
involving initial-state interactions shown in Fig.~\ref{fig2p}.  In
the following, we first discuss the role of initial-state
interactions, and argue that although required by gauge invariance,
they contribute to the $a$ term of Eq.~(\ref{BA13}) only.

After the collinear expansion 
of the parton momenta, the gluon interactions on the initial quark
lines, as shown in Fig.~\ref{fig2p}, can be reduced into two
categories: the long-distance and the short-distance contributions
due to the fact that every ``on-shell'' propagator can have a pole
contribution as well as a contact contribution \cite{Qiu}.  For
example, a quark propagator of momentum $k$ can always be written as  
\begin{equation}
\frac{i\gamma\cdot k}{k^2} 
= \frac{i\gamma\cdot \hat{k}}{k^2}
+ \frac{i\gamma\cdot n}{2k\cdot n}\ ,
\label{DF0}
\end{equation}
where $\hat{k}^2 = 0$ and $n^\mu$ is any auxiliary vector with $k\cdot
n \neq 0$.  The first term in the right-hand-side of Eq.~(\ref{DF0})
corresponds the pole contribution, while the second term is the
contact contribution \cite{Qiu}.  Attaching one gluon to the initial
quark line introduces a quark propagator, and this propagator will
have both the pole and contact contributions.  The pole contribution
is long-distance in nature, representing the interactions between the
quark and the gluon long before the hard collision between the quark
and the virtual photon.  The pole contribution is partially
responsible for the $A$-dependence of the leading-twist parton
distributions in a nucleus \cite{CQR}, and therefore, is one of the
sources for the weak $A$-dependence appeared in the $a$
term in Eq.~(\ref{BA13}). 

Because of the nature of the contact term, its contribution is
localized in space, and therefore, it does not contribute to the
$A^{1/3}$-type of nuclear enhancement \cite{QS_fac}.  Since the short
distance contributions of Feynman diagrams in Fig.~\ref{fig2p} have at
least one propagator given by the contact term, these diagrams do not
contribute to the $A^{1/3}$-type of nuclear enhancement.
At the same time, the contact term of the initial-state interaction
is important for the gauge invariance of the complete double
scattering (twist-4) process \cite{Qiu}, but only at order $A^0$. 

Although the Feynman diagrams with the final-state interactions
in Fig.~\ref{fig2} are not gauge invariant in general, their
contributions to the $A^{1/3}$-behavior of nuclear enhancement
are observable and hence gauge invariant \cite{LQS2}.  The three
lowest order diagrams in 
Fig.~\ref{fig2} are all convoluted with the same two-quark-two-gluon
matrix element through three independent parton momenta.  The leading
power contributions of these diagrams come from the region phase space
when all partons are collinear to the direction of the target, as 
shown in Fig.~\ref{fig2}.  In order to convert the gluon field $A^+$
to the corresponding field strength $F^{+\perp}$ in covariant gauge,
we expand the gluon momenta in an extra transverse component $k_T$ in 
Fig.~\ref{fig2} \cite{LQS2}.  All three Feynman diagrams
in Fig.~\ref{fig2} have two potential poles due to the interactions
between the final-state quark and gluons.  These two poles
(which are not pinched) can be used to carry out the momentum fraction
$dx_1$ and $dx_2$ integrals \cite{LQS1,LQS2}.  After taking the poles, 
the hard-scattering factor (the interaction between the
virtual photon and the initial quark) and the final-state interaction
between the ``on-shell'' outgoing quark and the gluon are separately
gauge invariant.

To derive the leading contribution from the double scattering, we
follow the derivation in Ref.~\cite{Guo2}.  By taking the poles to fix
the integrations of momentum fractions $x_1$ and $x_2$, we derive
contributions of all three diagrams in Fig.~\ref{fig2} 
to the hadronic tensor $W^{(D)}_{\mu\nu}$ in Eq.~(\ref{sigma-c}) as 
\begin{eqnarray}
W^{(D)}_{\mu\nu}(x_B,Q^2,p_c) &=& \frac{1}{4\pi}\,
\int \frac{dy^-}{4\pi}\, \frac{dy_1^-dy_2^-}{(2\pi)^2}\,
\frac{d^2y_T}{(2\pi)^2} \,
\int d^2 k_T\,  e^{ik_T \cdot y_{1T}} \, 
e^{i \frac{k_T^2-2k_T \cdot q}{2p \cdot q} p^+ (y_1^--y_2^-)} \,
\nonumber \\ 
&\ & \times\,
\langle p_A | \bar{\psi}_q(0)\,\gamma^+ \, 
A^+(y_{1}^{-},y_{1T})\,  A^+(y_{2}^{-})
            \psi_q(y^{-})\, | p_{A}\rangle 
\nonumber \\
&\ & \times (2\pi) \theta (y_1^--y^-) \, 
(2\pi) \theta (y_2^-) \, e^{i x_B p^+ y^- } 
\nonumber \\
&\ & \times \frac{2\pi}{(2p\cdot q)^3} 
      \cdot \hat{H}^{(D)}_{\mu\nu}(p_c) 
\cdot  \left[ \delta (p_{c_T}^2-k_T^2) - \delta (p_{c_T}^2) \right] 
dp_{c_T}^2 \\
& \approx & \left[
\frac{\pi}{(2p\cdot q)^3}\, \hat{H}^{(D)}_{\mu\nu}(p_c) \,
\left[-\delta ' (p_{c_T}^2) \right] T_{qF}^A(x_B,\mu^2) \right]
\, dp_{c_T}^2 \ .
\label{W-c}
\end{eqnarray}
In Eq.~(\ref{W-c}), $T_{qF}^A(x_B,\mu^2)$ is defined by
Eq.~(\ref{T4qA}) with the factorization scale $\mu^2$.  
$\hat{H}^{(D)}_{\mu\nu}(p_c)$ represents the spinor trace of the partonic
part of the double scattering and is given by 
\begin{equation}
\hat{H}^{(D)}_{\mu\nu}(p_c) =\frac{4}{3}
\pi^2 \alpha_s \alpha_{em} e_q^2 \,
{\rm Tr}\left[\gamma \cdot p \gamma_{\mu} \gamma \cdot (xp+q) 
\gamma^{\sigma} \gamma \cdot p_c \gamma^{\rho} \gamma \cdot (xp+q)
\gamma_{\nu}\right] \, p_{\rho}\,p_{\sigma}\ ,
\label{hatH}
\end{equation}
where a color factor $1/2N_c$ with $N_c=3$ was included.
Using the definition $z=\ell/p_c$, see Eq.~(\ref{z}) below, we can
reexpress the  
momentum $p_c$ in $\delta'(p^2_{c_T})$ in terms of the momentum $\ell$
of the observed pion,  
\begin{equation}
-\delta '(p_{c_T}^2)
=- \frac{d}{d(\frac{\ell_{T}^2}{z^2})} 
\left( \delta (\frac{\ell_{T}^2}{z^2}) \right) 
= - z^4 \delta '(\ell_{T}^2) \ .
\label{dpi}
\end{equation}
Combining Eqs.~(\ref{sigma-pi}), (\ref{sigma-c}), (\ref{W-c}),
and (\ref{dpi}), we obtain
\begin{equation}
\frac{d\sigma^{(D)}_{eA\rightarrow e\pi X}}{dx_B  dQ^2 d\ell_T^2}
=\frac{e^2}{8\pi x_B^2 s^2 Q^2} \, 
\frac{\pi}{(2p\cdot q)^3}
\sum_q\int dz\, z^2 \, D_{q\rightarrow \pi}(z)\, 
T_{qF}^A(x_B,Q^2)\, 
L^{\mu\nu} \hat{H}^{(D)}_{\mu\nu}(\ell/z) \,
\left[-\delta ' (\ell_{T}^2) \right]\, ,
\label{sigma-pi2}
\end{equation}
where the $\sum_q$ runs over all quark and antiquark flavors, and
$z=\ell/p_q$ is the momentum fraction of the quark flavor $q$ carried
by the observed pion.  At the lowest order in $\alpha_s$, hard gluon
initiated double scattering, which is proportional to the four-gluon
correlation function $T_{gF}$, does not contribute \cite{LQS1,LQS2}.

In the photon-nucleus frame, we choose the target momentum $p$
along the $-\vec{z}$-axis, such that
$p^{\mu}=(p^0,p^x,p^y,p^z)=(P,0,0,-P)$, 
and only $p^-=(p^0-p^z)/\sqrt{2}$ is nonvanishing.  In this frame, we 
have the following expression for the momentum fraction $z=\ell/p_c$, 
\begin{eqnarray}
z &\equiv& \frac{\ell}{p_{c}}=\frac{\ell^{+}}{p_{c}^{+}}
=\frac{p\cdot \ell }{p \cdot p_{c}} \nonumber \\ 
&\approx & \frac{p\cdot \ell }{p\cdot q} \ .
\label{z}
\end{eqnarray}
In deriving the last equation, we used $p_c=x_Bp+q$ and $p^2\approx
0$.   

After working out the algebra for $L^{\mu\nu} \hat{H}^{(D)}_{\mu\nu}$,
and substituting Eq.~(\ref{sigma-pi2}) into Eq.~(\ref{dlT20}), we
obtain   
\begin{equation}
\Delta\langle \ell_{T}^2\rangle_{1/3} \left(\ell_{\rm min}\right) = 
\frac{4\pi^2\alpha_s(\mu^2)}{3}\,
\frac{\sum_q e_q^2\,\int_{z_{\rm min}}^1 
             dz\, z^2 D_{q\rightarrow\pi}(z) T_{qF}^{A}(x_B,\mu^2)}
     {\sum_q e_q^2\,q^{A}(x_B,\mu^2)}\, ,
\label{pilt}
\end{equation}
where $D_{q\rightarrow\pi}(z)$ are the quark-to-pion fragmentation
functions, and $q^A(x_B,\mu^2)=\phi_{q/A}(x_B,\mu^2)$ are the
effective quark distributions in the nucleus normalized by the atomic
weight $A$.  In Eq.~(\ref{pilt}), $z_{\rm min}=p\cdot \ell_{\rm 
min}/(p\cdot q)$ is defined in terms of $\ell_{\rm min}$, which is 
the momentum cut for the pions measured in the experiment.  One can
choose $\ell_{\rm min}$ to be large enough to ensure that the observed 
pions are from the fragmentation of energetic quarks.

In Eq.~(\ref{pilt}), the $\mu^2$ in $\alpha_s(\mu^2)$
is the renormalization scale, and the $\mu^2$ in 
$T_{qF}^{A}(x_B,\mu^2)$ and $q^{A}(x_B,\mu^2)$ are the factorization
scale.  Since the short-distance part of the transverse momentum
broadening defined in this paper is an inclusive and infrared safe
calculable quantity, the scale $\mu^2$ should be chosen to be the
order of the physically measured momentum scales.  As shown in
Eq.~(\ref{pilt}), the only physically measured momentum scales for the 
transverse momentum broadening are $Q$ and $\ell_{\rm min}$.
Since $|\vec{\ell}_{\rm min}|$ are of the same order as
$\sqrt{Q^2}$, we choose $\mu^2=Q^2$ for the numerical calculations
below, and we believe that such a choice will not result into the
large logarithmic high order corrections.  

Since the quark-to-pion fragmentation functions have been measured
\cite{newDD}, Eq.~(\ref{pilt}) shows that the transverse momentum 
broadening of pions provides direct 
information on the quark-gluon correlation functions inside a 
nucleus.  The measured size of the transverse momentum broadening,
$\Delta\langle \ell_{T}^2\rangle_{1/3}$, depends on the choice of the
momentum cut $\ell_{\rm min}$ (or equivalently $z_{\rm min}$).  Our
predictions should be more reliable for the leading pions, or pions
with relatively large momenta.  By measuring the transverse momentum
broadening for $\pi^{\pm}$ and $\pi^0$, and keeping a reasonable large
value of $z_{\rm min}$, we can extract the quark flavor 
dependence of the correlation functions.


\section{Numerical Results and Discussions}

The numerical values of the pion transverse momentum broadening,
$\Delta\langle \ell_{T}^2\rangle_{1/3} (\ell_{\rm min})$, depend on the
explicit functional form of the quark-to-pion fragmentation functions
and the quark-gluon correlation functions.  In this section, without
assuming any specific form for the quark-gluon correlation functions,
we derive relations for the transverse momentum broadening between
$\pi^{\pm}$ and $\pi^0$ based on isospin 
symmetry and the  charge conjugation invariance.  Furthermore,
by using the simple model proposed in Ref.~\cite{LQS1} for the
quark-gluon correlation functions, we explore both the normalization
and functional dependence of the momentum broadening.

Although new parameterizations of the quark-to-pion fragmentation
functions were obtained recently \cite{newDD}, we will use the 
simple parameterizations of Ref.\cite{Dpi} for the following
analytical derivations and discussions on the general features of the 
momentum broadening.  Later, when we present our figures for 
the numerical values of the momentum broadening, we will use the 
parameterizations from both Refs.~\cite{Dpi} and \cite{newD}, and
demonstrate the similarities and differences.

Like the parton distributions, the quark-to-pion fragmentation
functions have scaling violation (or $Q^2$-dependence). To simplify
our discussion on the general features of transverse momentum
broadening, we ignore the scaling violation of the fragmentation
functions, and take for the fragmentation functions the input
distributions of Ref.~\cite{Dpi}, which are given as:
\begin{mathletters}
\label{D}
\begin{eqnarray}
D^+(z)=D_u^{\pi ^+}=D_d^{\pi ^-}=D_{\bar{u}}^{\pi ^-}
=D_{\bar{d}}^{\pi ^+} &=&\frac{1-z^2}{4z} \ ;
\label{D+} \\
D^-(z)=D_u^{\pi ^-}=D_d^{\pi ^+}=D_{\bar{u}}^{\pi ^+}
=D_{\bar{d}}^{\pi ^-} &=& \frac{(1-z)^2}{4z} \ ;
\label{D-} \\
D^0(z)=D_u^{\pi ^0}=D_d^{\pi ^0}=D_{\bar{u}}^{\pi ^0}
=D_{\bar{d}}^{\pi ^0} &=& \frac{1-z}{4z} \ ;
\label{D01} 
\end{eqnarray}
\end{mathletters}
with $D^0(z)=[D^+(z)+D^-(z)]/2$. 
For the strange quark fragmentation functions, we use  
$D_s^{\pi ^+}=D_s^{\pi ^-}=D_s^{\pi ^0}=D^-$.  
Notice that the fragmentation functions given in Eq.~(\ref{D}), as
well as those given in Ref.~\cite{newDD,newD}, violate Gribov-Lipatov  
reciprocity, which requires that the power of $(1-z)$ must be even.
However, since it is not our purpose to invent better fragmentation
functions in this paper, we will first use the fragmentation  
functions in Eq.~(\ref{D}) in our numerical calculation to illustrate 
the size and the general features of the transverse momentum
broadening.  Our predictions can be easily adjusted for other sets of
fragmentation functions. 

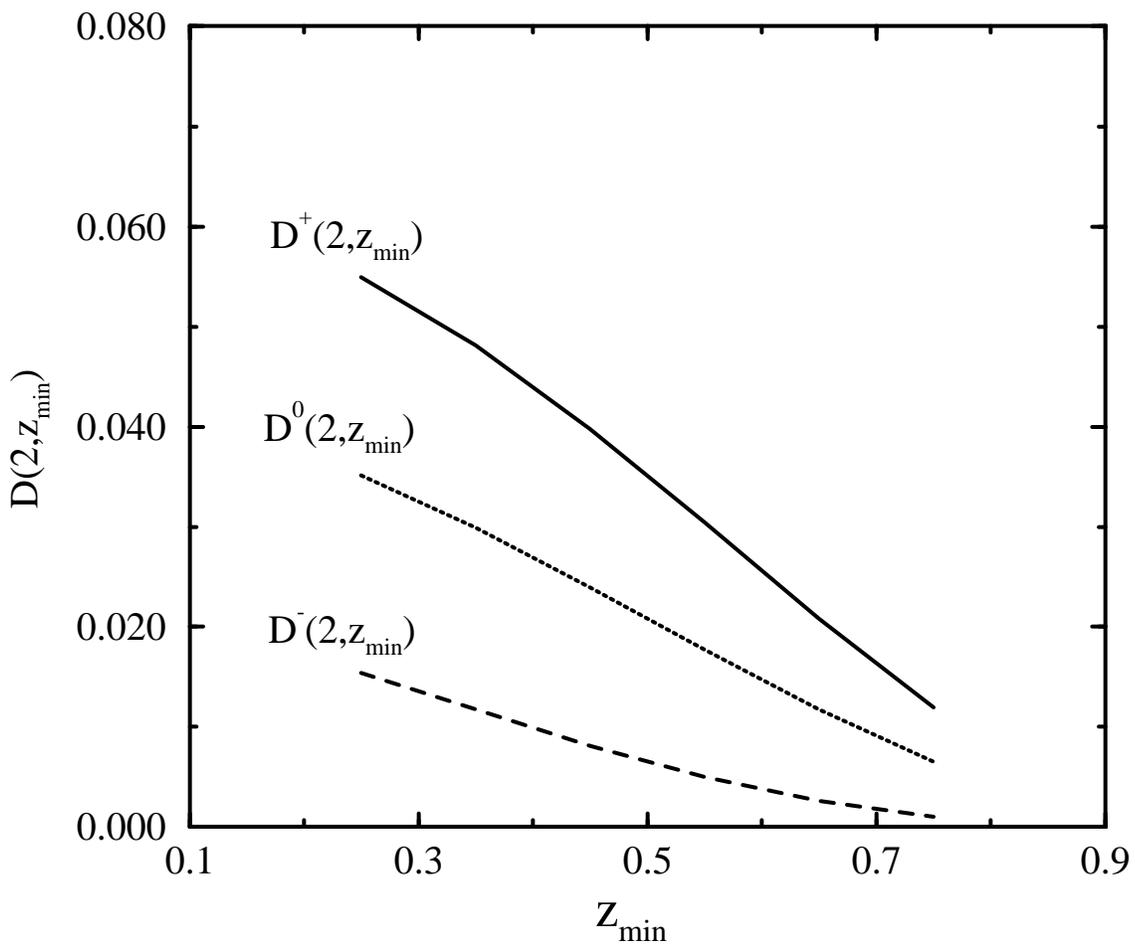
\begin{figure}
\caption{$D^+(2,z_{\rm min})$, $D^-(2,z_{\rm min})$, and 
$D^0(2,z_{\rm min})$ as functions of $z_{\rm min}$.}
\label{fig3}
\end{figure}

In order to evaluate
the transverse momentum broadening, Eq.~(\ref{pilt}) requires the
moments of the fragmentation functions.  We therefore introduce
\begin{equation}
D^{(i)}(n,z_{\rm min}) \equiv \int_{z_{\rm min}}^1 \,
dz\, z^n\, D^{(i)}(z) \ ,
\label{Dzn}
\end{equation}
where $i=+, -$ and $0$ for $\pi^+$, $\pi^-$ and $\pi^0$, respectively.
Using the fragmentation functions defined in Eq.~(\ref{D}), we obtain
the following identity,
\begin{equation}
D^{+}(n,z_{\rm min}) - D^{0}(n,z_{\rm min}) 
= D^{0}(n,z_{\rm min}) - D^{-}(n,z_{\rm min}) \ ,
\label{Dp0n}
\end{equation}
for all $n$.  This identity is useful for the following discussions on 
the general features of the transverse momentum broadening.  From
Eq.~(\ref{pilt}), the transverse momentum broadening, $\Delta\langle
\ell_{T}^2\rangle_{1/3} (\ell_{\rm min})$, 
depends on the second moments $D^{(i)}(2,z_{\rm
min})$, which are plotted in Fig.~\ref{fig3}.

Given the quark-to-pion fragmentation functions, we derive
the following general relations between the transverse momentum 
broadening of $\pi^+$, $\pi^-$, and $\pi^0$ particles:
\begin{mathletters}
\begin{equation}  
\Delta \langle \ell_{T}^2\rangle_{1/3}^{\pi^+}
- \Delta \langle \ell_{T}^2\rangle_{1/3}^{\pi^0} =
\Delta \langle \ell_{T}^2\rangle_{1/3}^{\pi^0}
- \Delta \langle \ell_{T}^2\rangle_{1/3}^{\pi^-} \ ,
\label{symmetry}
\end{equation}
and
\begin{equation}
\Delta \langle \ell_{T}^2\rangle_{1/3}^{\pi^+}
> \Delta \langle \ell_{T}^2\rangle_{1/3}^{\pi^0} 
> \Delta \langle \ell_{T}^2\rangle_{1/3}^{\pi^-} \ ,
\label{p0m}
\end{equation}
\end{mathletters}
independent of the details of the correlation functions $T_{qF}^A$.
Eq.~(\ref{symmetry}) is a result of the isospin symmetry of the
fragmentation functions given in Eq.~(\ref{D}) and the identity 
given in Eq.~(\ref{Dp0n}). The inequality given in Eq.~(\ref{p0m}) 
results from the fact that the contribution of each quark flavor is 
weighted by the square of the quark's fractional charge, $e_q^2$, 
and the relation $D^+(z)>D^0(z)>D^-(z)$. 

Since the correlation functions $T_{qF}^A$ are not known, 
in order to obtain numerical estimates of  the transverse
momentum broadening of pions, we adopt the following model 
for quark-gluon correlation functions \cite{LQS1,LQS2}:
\begin{equation}
T_{qF}^A(x,Q^2)=\lambda^2 A^{1/3} q^A(x,Q^2) \ ,
\label{TiM}
\end{equation}
where $q^A(x)$ is the effective twist-2 quark distribution of a 
nucleus normalized by the atomic weight $A$, and $\lambda$ is a free
parameter to be fixed by experimental data.  This model was proposed 
after comparing the operator definitions of the four-parton 
correlation functions and the definitions of the normal 
twist-2 parton distributions.  Substituting Eq.~(\ref{TiM}) into 
Eq.~(\ref{pilt}), we have 
\begin{equation}
\Delta\langle \ell_{T}^2\rangle_{1/3} = 
\frac{4\pi^2\alpha_s(Q^2)}{3}\, A^{1/3}\, \lambda^2\,
\frac{\sum_q e_q^2\,q^{A}(x_B,Q^2)\, 
      D_{q \rightarrow \pi}(2,z_{\rm min})}
     {\sum_q e_q^2\,q^{A}(x_B,Q^2)}\, .
\label{pilt2}
\end{equation}

From Eq.~(\ref{pilt2}),  we can deduce a very simple formula for 
the transverse momentum broadening of $\pi^0$ by using the
fragmentation functions given in Eq.~(\ref{D01}):
\begin{equation}  
\Delta \langle \ell_{T}^2\rangle_{1/3}^{\pi^0} 
\approx \frac{4\pi^2\alpha_s(Q^2)}{3}\, A^{1/3}\, \lambda^2\,
D^0(2,z_{\rm min})\ .
\label{ltpi0}
\end{equation}
In deriving Eq.~(\ref{ltpi0}), we neglected the strange quark
contribution which is much smaller. 
This simple expression shows that 
$\Delta \langle \ell_{T}^2\rangle_{1/3}^{\pi^0}$ has a scaling 
behavior as $x_B$ varies.  The approximate $x_B$-scaling behavior 
of $\Delta \langle \ell_{T}^2\rangle_{1/3}^{\pi^0}$ is a direct 
consequence of the model for $T_{qF}^A(x)$, 
given in Eq.~(\ref{TiM}).  If the future experimental data on the
transverse momentum broadening of $\pi^0$ shows a strong violation of
the $x_B$-scaling, the model for the quark-gluon correlation function
$T_{qF}^A(x_B)$, given in Eq.~(\ref{TiM}), will have to be modified.  
In addition, Eq.~(\ref{ltpi0}) also shows that  $Q^2$ dependence 
of $\Delta \langle \ell_{T}^2\rangle_{1/3}^{\pi^0}$ is mainly from the
$Q^2$ dependence of $\alpha_s(Q^2)$. Since the $Q^2$ dependence 
of $\alpha_s(Q^2)$ is already known, we can learn the $Q^2$ 
dependence of the correlation function $T_{qF}^A(x_B,Q^2)$ 
by measuring the $Q^2$ dependence of 
$\Delta \langle \ell_{T}^2\rangle_{1/3}^{\pi^0}$. If the measured 
$Q^2$ dependence of  $\Delta \langle \ell_{T}^2\rangle_{1/3}^{\pi^0}$ 
is consistent with the $Q^2$ dependence of $\alpha_s(Q^2)$, 
it means the correlation function $T_{qF}^A(x_B,Q^2)$ has a
similar scale-dependence as a normal parton distribution. 
By examining the $x_B$-scaling property and $Q^2$ dependence
of $\Delta \langle \ell_{T}^2\rangle_{1/3}^{\pi^0}$, we can 
test whether the model for $T_{qF}^A(x_B,Q^2)$, given in 
Eq.~(\ref{TiM}), is reasonable.

It should be emphasized that our general conclusions given in last
paragraph, as well as numerical estimates given below, are the
consequences of our lowest order calculations.  High order corrections
in $\alpha_s$ and/or in inverse powers of $Q^2$ can certainly
change the dependence on $x_B$ and $Q^2$, as well as the absolute
values of our numerical estimates.  But, because of the inclusivity
and the infrared safety of the transverse momentum broadening, and the
lack of the large logarithms of the ratios of the physically measured
scales, we believe that the change should not be very dramatic, and
that our predictions should not be off by orders of magnitudes.

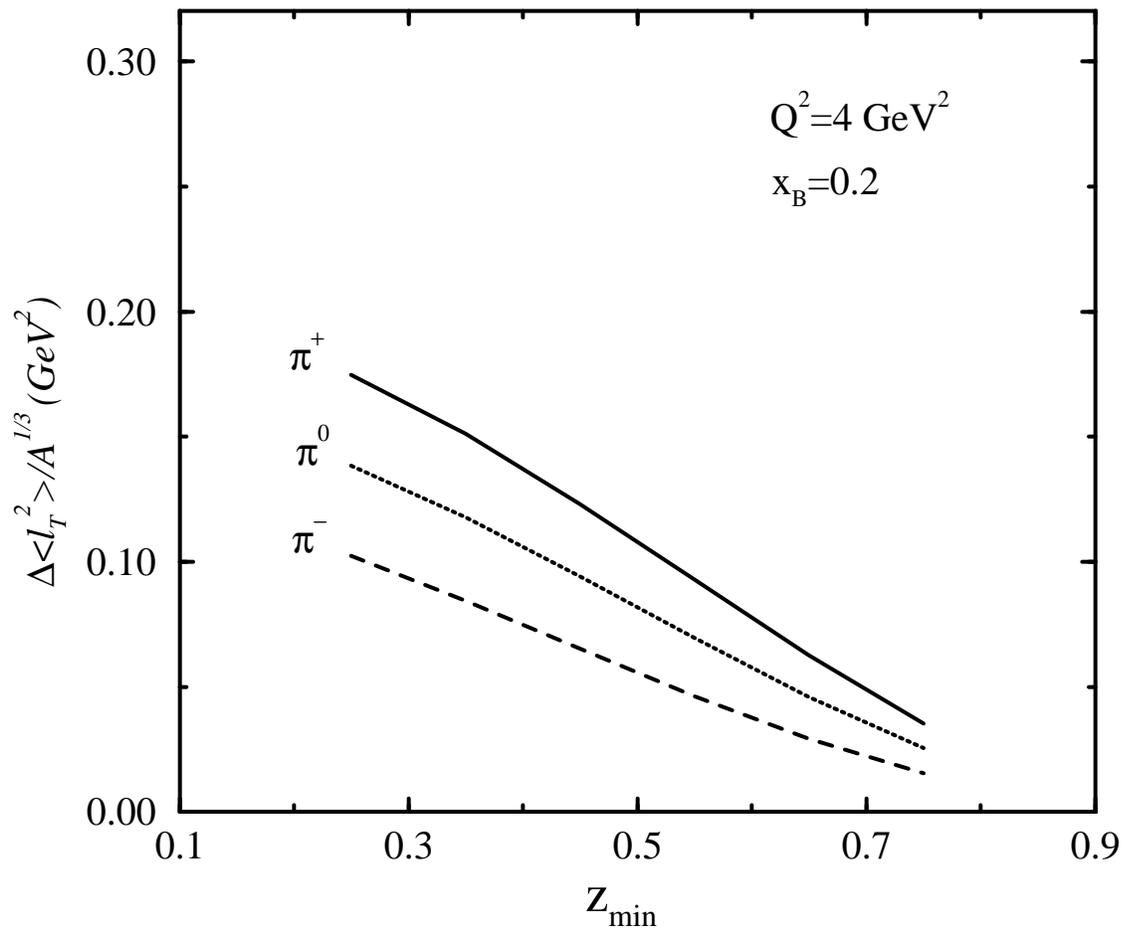
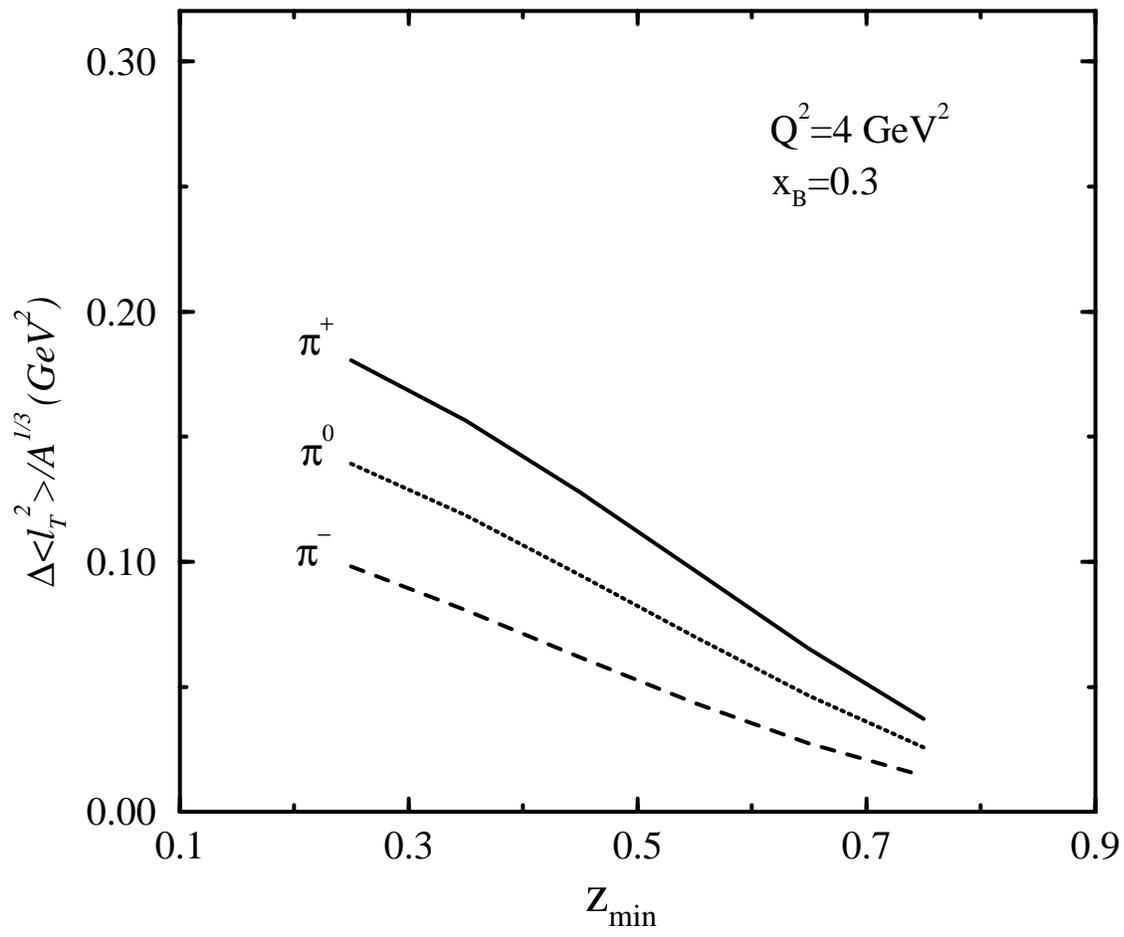
\begin{figure}
\caption{ Transverse momentum broadening of pions, 
 $\Delta\langle \ell_{T}^2\rangle_{1/3} /A^{1/3}$, 
as a function of $z_{\rm min}$ at $Q^2=4$ GeV$^2$ and   
(a) $x_B=0.2$, and (b) $x_B=0.3$. }
\label{fig5}
\end{figure}

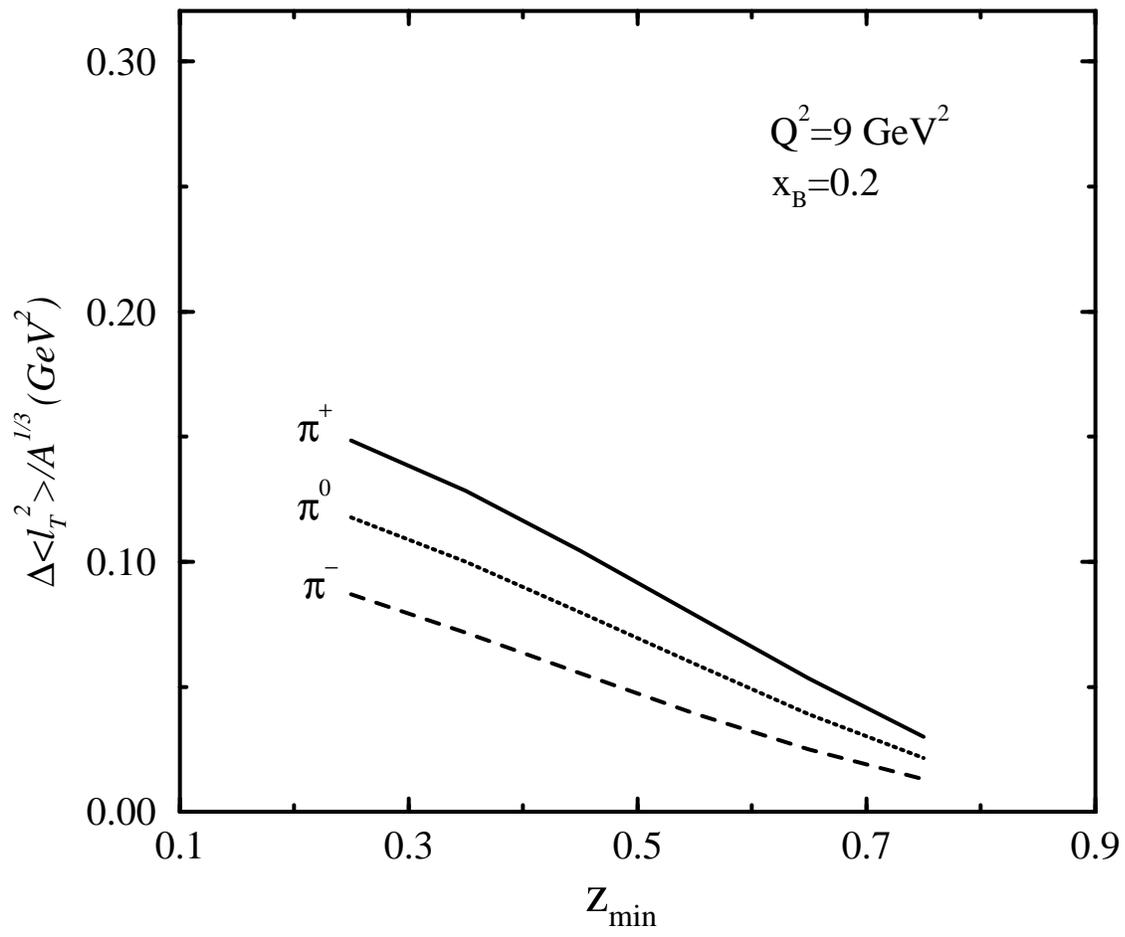
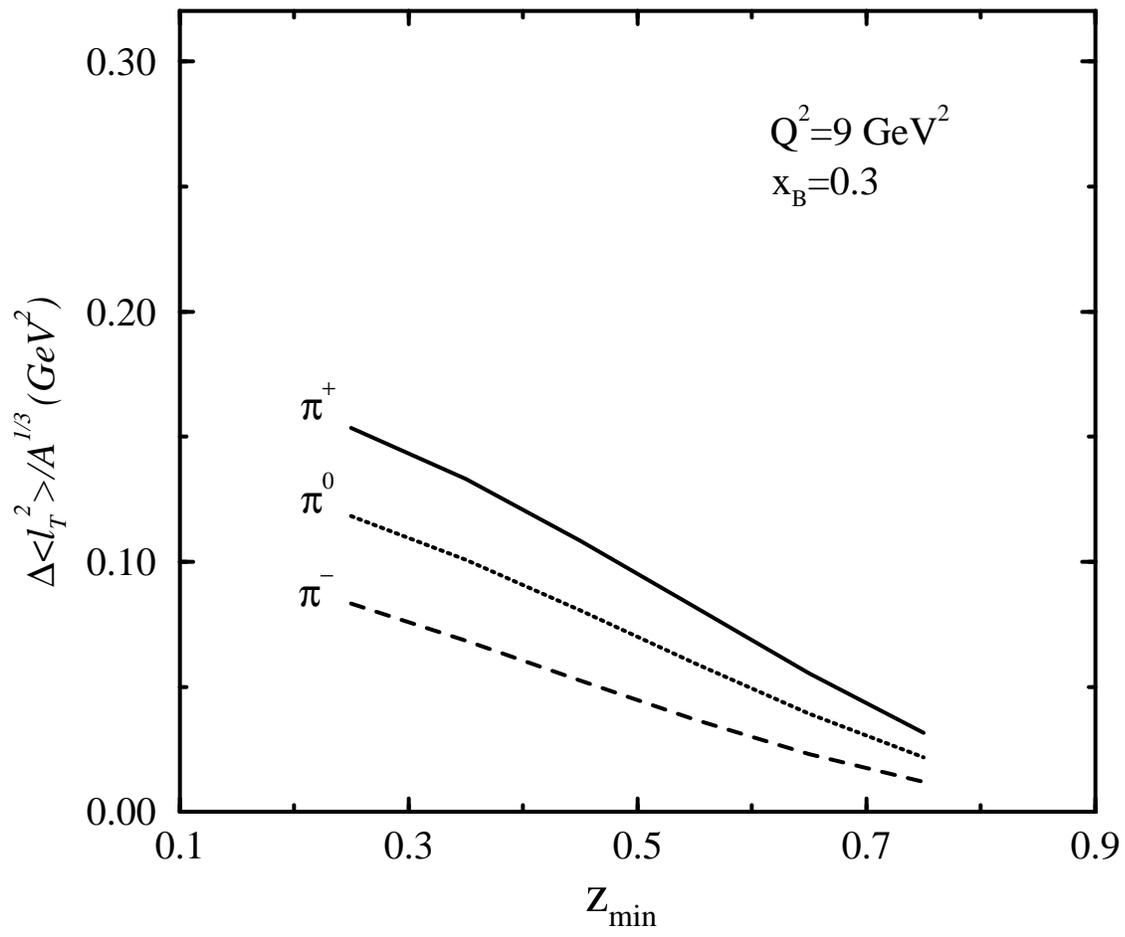
\begin{figure}
\caption{ Transverse momentum broadening of pions,
$\Delta\langle \ell_{T}^2\rangle_{1/3} /A^{1/3}$,  
as a function of $z_{\rm min}$ at $Q^2=9$ GeV$^2$ and 
(a) $x_B=0.2$, and (b) $x_B=0.3$.}
\label{fig6}
\end{figure}

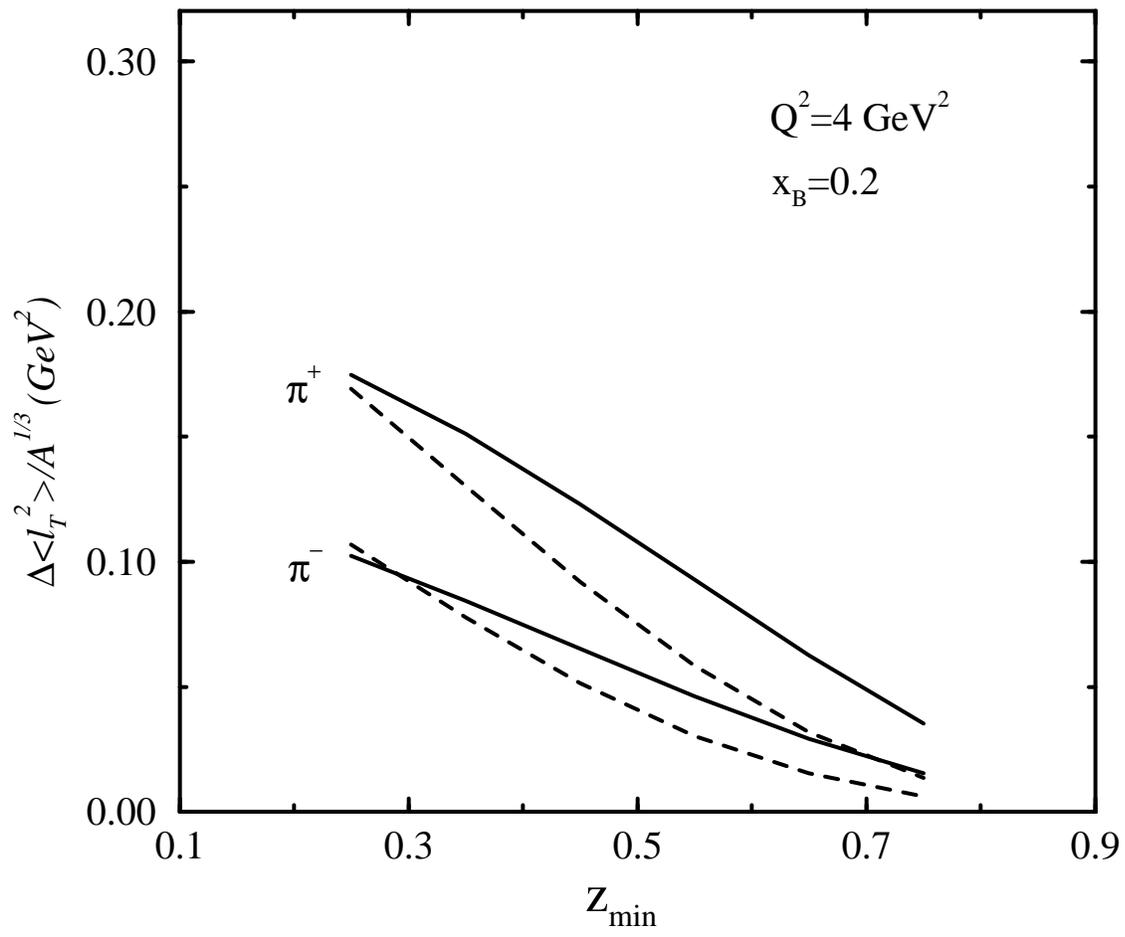
\begin{figure}
\caption{
Transverse momentum broadening of pions,
$ \Delta \langle l_T^2 \rangle_{1/3}^{\pi^{\pm} } / A^{1/3} $, at 
$Q^2=4$ GeV$^2$ and $x_B=0.2$ with different 
$D_{q\rightarrow \pi}(z)$.  The solid lines are
obtained by using the fragmentation functions of 
Ref.~\protect\cite{Dpi}, and the dashed lines are obtained by 
using the fragmentation functions of Ref.~\protect\cite{newD}.}
\label{fig7}
\end{figure}

In the following, we obtain our numerical estimates of 
transverse momentum broadening for $\pi^+$, $\pi^-$, and 
$\pi^0$ particles by evaluating Eq.~(\ref{pilt2}).  We normalize     
$\Delta\langle \ell_{T}^2\rangle_{1/3}$ by $A^{1/3}$, and plot our 
results in Fig.~\ref{fig5} to Fig.~\ref{fig6}.  
In view of the limited range of $Q^2$ available at CEBAF, 
we neglect the scaling violation of the fragmentation functions 
for our numerical estimates. We used the simple fragmentation 
functions given  in Eq.~(\ref{D}) to produce 
Fig.~\ref{fig5} to Fig.~\ref{fig6}.  More realistic  
parameterizations for quark-to-pion  fragmentation 
functions were obtained in Refs.~\cite{newDD} and \cite{newD}.  In
particular, those in Ref.~\cite{newD} are extracted from DIS data.  
We present both results for 
$\Delta\langle \ell_{T}^2\rangle_{1/3}^{\pi^\pm} / A^{1/3}$ 
by using two different sets of the fragmentation functions at
$Q^2=4$~GeV$^2$ in Fig.~\ref{fig7} to illustrate the uncertainties in
our numerical estimates due to different choices of the fragmentation  
functions.  Although numerical details of the $A^{1/3}$-type
transverse momentum broadening depend on the fragmentation functions
used, the overall shape and magnitude of the broadening are consistent.
The most important feature is that the size of the $A^{1/3}$-type
broadening is large enough to be measured experimentally
\cite{E683,E609}. 

In obtaining our numerical results, we used CTEQ3L parton 
distributions of Ref.~\cite{CTEQ3} for the quark distributions 
in the nucleons. We took an average of the quark distributions of  
the proton and the neutron for the effective quark distribution 
in the nucleus, $q^A(x_B)$.  If the $A$-dependence of the nuclear
quark distributions is independent of the flavor,  
from Eq.~(\ref{pilt2}), such nuclear dependence can be factorized 
and canceled between the numerator and denominator.
However, there is no obvious reason why the nuclear dependence of
quark distributions is flavor independent.  But, as we discussed
above, in the range of $x$ probed in these experiments the effective
nuclear parton distribution should have  
extremely weak nuclear dependence in comparison with the
$A^{1/3}$-type enhancement that we discussed in this paper.  
In order to verify this feature, we used both $q^A(x_B)=q^N(x_B)$ and
the realistic parameterizations for $q^A(x_B)$ given in
Ref.~\cite{Eskola} to calculate the transverse momentum broadening 
$\Delta\langle \ell_{T}^2\rangle_{1/3}^{\pi^\pm} / A^{1/3}$.  
We found no noticeable difference for the 
transverse momentum broadening presented in Figs.~\ref{fig5},
\ref{fig6}, and \ref{fig7}, which were obtained with 
$q^A(x_B)=q^N(x_B)$.

Figs.~\ref{fig5}a and \ref{fig5}b shows 
$\Delta \langle \ell_{T}^2\rangle_{1/3} / A^{1/3}$ 
as functions of $z_{\rm min}$ at $Q^2=4$ GeV$^2$, and 
$x_B=0.2$ and $0.3$, respectively.
Figs.~\ref{fig6}a and \ref{fig6}b show  
$\Delta \langle \ell_{T}^2\rangle_{1/3} / A^{1/3}$ as functions 
of $z_{\rm min}$ at $Q^2=9$ GeV$^2$, and $x_B=0.2$ and
$0.3$, respectively.  In these figures, the relations 
$\Delta \langle \ell_{T}^2\rangle_{1/3}^{\pi^+}
> \Delta \langle \ell_{T}^2\rangle_{1/3}^{\pi^0} 
> \Delta \langle \ell_{T}^2\rangle_{1/3}^{\pi^-}$
and 
$\Delta \langle \ell_{T}^2\rangle_{1/3}^{\pi^+}
- \Delta \langle \ell_{T}^2\rangle_{1/3}^{\pi^0} =
\Delta \langle \ell_{T}^2\rangle_{1/3}^{\pi^0}
- \Delta \langle \ell_{T}^2\rangle_{1/3}^{\pi^-}$
are clearly demonstrated. Furthermore, we have $x_B$-scaling for 
$\Delta \langle \ell_{T}^2\rangle_{1/3}^{\pi^0}$, and 
$\Delta \langle \ell_{T}^2\rangle_{1/3}^{\pi^+} / 
\Delta \langle \ell_{T}^2\rangle_{1/3}^{\pi^-} \approx 2$
for small $z_{\rm min}$. The approximate ratio,
$\Delta \langle \ell_{T}^2\rangle_{1/3}^{\pi^+} / 
\Delta \langle \ell_{T}^2\rangle_{1/3}^{\pi^-} \approx 2$,
is a result of the isospin averaged targets, and 
can be easily verified by using the
fragmentation functions in Eq.~(\ref{D}), keeping   
only the valence quark contributions and 
$z_{\rm min}^2 \ll 1$.

Comparing Figs.~\ref{fig5} and \ref{fig6}, 
it is evident that the absolute sizes of the 
$\Delta \langle \ell_{T}^2\rangle_{1/3}$ decrease as $Q^2$ 
increases.  This is caused by the $Q^2$ dependence of the 
$\alpha_s(Q^2)$ in the overall factor of the transverse 
momentum broadening.  Due to the available energy at CEBAF, we plotted
the broadening at $Q^2=4$~GeV$^2$ in Fig.~\ref{fig5}.  Even though   
$Q^2=4$~GeV$^2$ may not be large enough to apply for the fragmentation 
analysis, our results can be tested at CEBAF with its future upgrade,
and can be easily tested at RHIC with its future option of
electron-ion collider.

From Figs.~\ref{fig5} and \ref{fig6}, we noticed that the identity, 
$\Delta \langle \ell_{T}^2\rangle_{1/3}^{\pi^+}
- \Delta \langle \ell_{T}^2\rangle_{1/3}^{\pi^0} = 
\Delta \langle \ell_{T}^2\rangle_{1/3}^{\pi^0}
- \Delta \langle \ell_{T}^2\rangle_{1/3}^{\pi^-}$,
has some weak violation on the value of $x_B$, which is  
due to the fact that the number of sea quark increases 
at smaller $x_B$ in comparison with the valence quarks.

The value of $\lambda^2$ for the quark-gluon correlation 
function $T_{qF}^A$, given in Eq.~(\ref{TiM}), has not 
been well determined. 
It was estimated in Ref.~\cite{LQS2} by using the
\underline{measured} nuclear enhancement of the momentum imbalance of
two jets in photon-nucleus collisions \cite{E683,E609}, and was found 
to be the order of $\lambda^{2} \sim 0.05 - 0.1 \mbox{GeV}^{2}$.
However, the nuclear enhancement of the Drell-Yan transverse momentum 
\cite{NA10,E772} indicates a much smaller value of $\lambda^2$ 
\cite{Guo2}.  Although the momentum imbalance of the di-jet data 
depends on the final-state multiple scattering, and the Drell-Yan
data is an effect of initial-state multiple scattering, the leading 
order calculation indicates that $\lambda^2$ from
the two data sets to be the same \cite{Guo2}.  Therefore,
this work on the transverse momentum broadening in pion productions
should provide a new and independent measurement of the size of the 
quark-gluon correlation functions, and the value of the $\lambda^2$,
which is extremely important for understanding the nuclear dependence 
and the event rates in the relativistic heavy ion collisions.
Since both di-jet momentum imbalance and the transverse momentum 
broadening are caused by the final state multiple scattering, we used 
$\lambda^2=0.05$~GeV$^2$ for Figs.~\ref{fig5} to \ref{fig7}. 
Any change in $\lambda^2$ will only change the overall normalization 
factor in $\Delta\langle \ell_{T}^2\rangle_{1/3} /A^{1/3}$, and hence 
does not affect the general features of 
$\Delta\langle \ell_{T}^2\rangle_{1/3} /A^{1/3}$.

We conclude that the transverse momentum broadening
of leading pions in deeply inelastic lepton-nucleus scattering 
is an excellent observable to probe the parton correlation 
functions in the nucleus.  Measuring the $A^{1/3}$-type
transverse momentum broadening $\Delta\langle \ell_{T}^2\rangle_{1/3}$
in the leading pion production in DIS provides an independent test for
the existing model of the quark-gluon correlation functions
\cite{LQS1}.  More importantly, by measuring such broadening, we can  
directly measure the $x$ and $Q^2$ dependence of the correlation
functions, which is very useful for predicting the event rates
\cite{FMSS} and for understanding the nuclear dependence in 
relativistic heavy ion collisions.  


\section*{Acknowledgment}

We thank G. Sterman for very helpful discussions. 
This work was supported in part by the U.S. Department of Energy under
Contract No. DE-AC02-98CH10886, and Grant Nos. DE-FG02-87ER40731 and  
DE-FG02-96ER40989.    


\end{document}